\documentclass[aps,preprint,prd,nofootinbib,epsfig]{revtex4-1}
\pdfoutput=1 
\usepackage{pifont}
\usepackage{color}
\usepackage{amssymb}
\usepackage{amsmath}
\usepackage{graphicx}
\usepackage{fancyhdr}
\newcommand{\be}{\begin{equation}}
\newcommand{\ee}{\end{equation}}
\newcommand{\bea}{\begin{eqnarray}}
\newcommand{\eea}{\end{eqnarray}}
\newcommand{\bes}{\begin{subequations}}
\newcommand{\ees}{\end{subequations}}

\newcommand{\bc}{\begin{center}}
\newcommand{\ec}{\end{center}}
\begin{document}
\title{  Higgs mass and right-handed sneutrino WIMP in a  supersymmetric 3-3-1 model}
\author{C. A. de S. Pires, P. S. Rodrigues da Silva, A. C. O. Santos, Clarissa Siqueira}
\affiliation{ Departamento de F\'{\i}sica, Universidade Federal da Para\'\i ba, Caixa Postal 5008, 58051-970,
Jo\~ao Pessoa, PB, Brazil}

\date{\today}

\begin{abstract}
This work deals with right handed sneutrino as thermal cold dark matter candidate. This scalar emerges in a supersymmetric version of $SU(3)_c \otimes SU(3)_L \otimes U(1)_X$ gauge model where right handed neutrinos are a natural component of leptonic chiral scalar supermultiplets. We first consider the issue of a $125$~GeV Higgs boson mass in this model, showing that constraints on stop mass and trilinear soft coupling are considerably alleviated compared to MSSM. Then we investigate the region of parameter space that is consistent with right handed sneutrino as thermal cold dark matter, under the light of Planck results on the relic abundance and direct detection from LUX experiment. This sneutrino mainly annihilates through an extra neutral gauge boson, $Z^\prime$, and Higgs exchange, so that the physics of dark matter is somewhat related to the parameters determining Higgs and $Z^\prime$ masses.
We then obtain that right handed sneutrino in this model must be heavier than $400$~GeV to conform with Planck and LUX, simultaneously constraining the $Z^\prime$ mass to be above 2400~GeV, which is in perfect agreement with LHC searches in a non-supersymmetric version of this model.
\end{abstract}

\maketitle

\section{Introduction}
\label{intro}
The amount of missing mass in the universe, the so called dark matter, has been precisely determined after WMAP~\cite{Hinshaw:2012aka} and Planck~\cite{Ade:2015xua} satellites. However, there seems to be no real appealing solution to this problem besides it being constituted of new neutral and stable particle(s) beyond those already known. Several facilities were aimed to detect it directly~\cite{Agnese:2013rvf,Aprile:2012nq,Reindl:2015dnp,Akerib:2015rjg}, mainly when it lies in the range of hundreds of GeV mass scale, characterizing what is known as a weakly interacting massive particle (WIMP)~\cite{Jungman:1995df,Bertone:2004pz}. The WIMP paradigm is so largely accepted because it miraculously fits to what is expected from a natural extension of the standard model of electroweak interactions (SM), realized close to its symmetry breaking scale, around 1~TeV, and whose interactions are sort of weak too, allowing for the observed abundance of cold dark matter (CDM). 
Concomitantly, there are strong reasons to believe that Supersymmetry (SUSY) may exist at very high energies and be broken close to  the electroweak scale, being phenomenologically accessible at the Large Hadron Collider (LHC).
If SUSY is armed with R-parity symmetry for the component fields, $R=(-1)^{3(B-L)+2s}$, a discrete symmetry that may be a remnant of a $U(1)_{B-L}$ lepton-baryon number gauge symmetry, avoiding the proton decay, it simultaneously provides a stable supersymmetric particle with the right features to be a WIMP.

Among the neutral supersymmetric particles, sneutrino~\cite{Hagelin:1984wv,Ibanez:1983kw,Falk:1994es}  as well as neutralino~\cite{Jungman:1995df} are the two kinds of particles that may play the role of WIMPs. However, in the minimal supersymmetric standard model (MSSM), only  neutralino is  viable as CDM candidate because the left-handed sneutrino  has a sizable coupling with the $Z^0$ boson and, consequently,  either gives a small relic abundance or is excluded by direct CDM  searches~\cite{Agnese:2013rvf,Aprile:2012nq,Reindl:2015dnp,Akerib:2015rjg}.  
It would be interesting to look for extensions of the MSSM that could accommodate both forms of WIMPs as viable CDM (not simultaneously though), augmenting the chances of describing it while conforming with phenomenological constraints over the model.  In this direction, there is no other alternative  unless to consider the scalar superpartner of the  right-handed neutrino
~\cite{Asaka:2005cn,Gopalakrishna:2006kr,Asaka:2006fs,
Lee:2007mt,Arina:2007tm,Cerdeno:2008ep,Cerdeno:2009dv,Belanger:2010cd,Belanger:2011rs,Dumont:2012ee,
DeRomeri:2012qd,Tang:2016oba}.

Instead of only adding a new singlet superfield to the MSSM to obtain the right handed sneutrino, we call on the  supersymmetric version of a gauge extension of the SM, the  $SU(3)_C \times SU(3)_L \times U(1)_X$ (3-3-1) gauge model that already possesses right-handed neutrinos as a natural ingredient of their particle content~\cite{Singer:1980sw,Foot:1994ym,Montero:1992jk}.
This class of model presents appealing features, one of them being the fact that a minimal of tree families are necessary in order to cancel anomalies, offering an explanation to the old family puzzle~\cite{Pisano:1991ee,Frampton:1992wt}. They also shed some light on the understanding of the  quantization of electric charges~\cite{deSousaPires:1998jc} and provide a solution to the strong CP  problem~\cite{Pal:1994ba,Dias:2003zt,Dias:2003iq}, address the neutrino mass and oscillation pattern~\cite{Gusso:2003cp,Dias:2005yh,Cogollo:2008zc,Queiroz:2010rj,Pires:2014xsa}, possesses neutral stuff that can be accommodated in a WIMP framework~\cite{deS.Pires:2007gi,Mizukoshi:2010ky,Alvares:2012qv,daSilva:2014qba}, can account for a possible extra radiation imprint on the Cosmic Microwave Background Radiation~\cite{Kelso:2013nwa,Queiroz:2013lca}, among others. These features surely confer enough motivation that justifies the development of such class of gauge models and their supersymmetric versions~\footnote {Although  3-3-1 gauge models are becoming popular and very well developed, their supersymmetric versions have received scarce attention. For some works considering SUSY 3-3-1, see Ref.~\cite{Montero:2000ng,Montero:2004uy,Dong:2006vk,Dong:2007qc}} .

In this work we study the Higgs and the  dark matter sector of the supersymmetric version of the 3-3-1 model with right-handed neutrinos (S331RH$\nu$). We argue that right-handed sneutrino, decoupled from the $Z^0$ gauge boson,  is the simplest form of CDM candidate provided by the model. We then  calculate its relic abundance and investigate the direct detection of  sneutrino as WIMP. In order to be sure that our results are realistic, we also investigate the scalar sector of the model and show that a Higgs with a mass of $125$ GeV with stop mass and soft trilinear coupling below TeV scale is a  natural outcome of the model.

The paper is divided in the following way: In the first section, sec.~\ref{sec1}, we introduce the main ingredients of the model, identifying its content, mass spectrum, superpotential and soft SUSY breaking terms according to the gauge and discrete symmetries imposed. Next, in sec.~\ref{sec2}, we focus on numerical calculation of the Higgs mass in this model, looking at the leading quantum contribution. We then, in Sec.~\ref{sec3}, analyze the sneutrino as a WIMP candidate by computing its relic abundance and direct detection cross section, contrasting them with observation. We finally conclude in Sec.~\ref{sec4}.

\section{The main ingredients of the model}
\label{sec1}

In the leptonic sector, the superfields of the three generations compose triplet and singlet representations according to the following transformation by the 3-3-1 symmetry, 
\begin{eqnarray}
\hat L_{a} = \left (
\begin{array}{c}
\hat \nu_{a} \\
\hat e_{a} \\
\hat \nu^c_{a}
\end{array}
\right )_L\sim(1\,,\,3\,,\,-1/3)\,,\,\,\, \hat l^c_{aL}\,\sim(1,1,-1)\,,
\label{Lcontent}
\end{eqnarray}
where $a=1,2,3$ represents the family index for the usual three generations of leptons. Observe that  right-handed neutrinos are incorporated as the third component of a fundamental representation of $SU(3)_L$ for leptons, while the right-handed charged leptons are singlets under this symmetry.. 

In the Hadronic sector, the superfields of the  third generation come in the triplet representation and the superfields of the other two are in anti-triplet representations of $SU(3)_L$, as a requirement for anomaly cancellation. They are given by,
\begin{eqnarray}
&& \hat{Q}_{\alpha_L}=\left(\begin{array}{c}
                 \hat{d}_{\alpha}  \\
                 \hat{u}_{\alpha}  \\
                 \hat{d}^{\prime}_{\alpha}
                 \end{array}\right)_L \sim (3,3^*,0) ; \nonumber \\
                 && \hat u_{\alpha_L}^{\,c} \sim(3^*,1,-2/3)\,,\,\,\, \hat d_{\alpha_L}^{\,c} ,\hat d^{\prime{\,c}}_{\alpha_L} \sim(3^*,1,1/3)\,,\,\,\, \nonumber \\
 &&\hat{Q}_{3 L}=\left(\begin{array}{c}
                 \hat{u}_3  \\
                 \hat{d}_3  \\
                 \hat{u}^{\prime}_3 
                \end{array}\right)_L \sim (3,3,\frac{1}{3}), \nonumber \\
                &&
  \hat u_{3L}^c, \hat u_{3L}^{\prime c}\sim(3^*,1,-2/3)\,,\,\,\, \hat d_{3L}^{\,c}\sim(3^*,1,1/3)
\label{Hcontent}
\end{eqnarray}
where $\alpha = 1,2$.

The scalars of the model, responsible for the spontaneously broken gauge symmetry, compose the following superfields, 
\begin{eqnarray}
 \label{part_transf1}
 \hat{\eta}=\left(\begin{array}{c}
                 \hat{\eta}_1  \\
                 \hat{\eta}^{-}  \\
                 \hat{\eta}_2 \\
                 
\end{array}\right) , 
 \hat{\chi}=\left(\begin{array}{c}
                 \hat{\chi}_1  \\
                 \hat{\chi}^{-}  \\
                 \hat{\chi}_2 \\
                 
\end{array}\right), 
 \hat{\rho}=\left(\begin{array}{c}
                 \hat{\rho}_1^{+}  \\
                 \hat{\rho}  \\
                 \hat{\rho}_2^{+} \\
                 
\end{array}\right),
\end{eqnarray}
where $\hat{\eta}, \hat{\chi}\sim (1,3,-1/3)\,,\,\,\,$  $\hat{\eta}\sim (1,3,2/3)$, and,
\begin{eqnarray}
 \label{part_transf2}
 \hat{\eta}^{\prime}=\left(\begin{array}{c}
                 \hat{\eta}^{\prime}_1  \\
                 \hat{\eta}^{\prime -}  \\
                 \hat{\eta}^{\prime}_2 \\
                 
\end{array}\right) , 
 \hat{\chi^{\prime}}=\left(\begin{array}{c}
                 \hat{\chi}^{\prime}_1  \\
                 \hat{\chi}^{\prime -}  \\
                 \hat{\chi}^{\prime}_2 \\
                 
\end{array}\right), 
 \hat{\rho}^{\prime}=\left(\begin{array}{c}
                 \hat{\rho}^{\prime}_1\!\!\! .^{+}  \\
                 \hat{\rho}^{\prime} \\
                 \hat{\rho}^{\prime}_2\!\!\! .^{+} \\
                 
\end{array}\right),
\end{eqnarray}
where $\hat{\eta}^{\prime}, \hat{\chi}^{\prime}\sim (1,3^*,1/3)\,,\,\,\,$ $\hat{\rho}^{\prime}\sim (1,3^*,-2/3)$. 
It is opportune to remark that the nonsupersymmetric version of this model demands a total of at least three scalar triplets in order to engender spontaneous symmetry breaking and describe fermion masses. The scalars that transform in the same way  ($\hat{\eta}$ and $ \hat{\chi}$, for example)  have different neutral components developing vacuum expectation value (VEV) in a way that lepton number is conserved by the vacuum. This is the reason to have two such triplets.
Considering this and given their quantum numbers, we are obliged to duplicate all the three scalar triplets associating opposite quantum numbers to them so as to cancel gauge anomalies, justifying the choice above.

For reasons of simplicity (and avoiding spontaneous lepton number violation), we assume that only  the neutral scalars  $\eta_1$, $\eta^{\prime}_1$,  $\rho$, $\rho^{\prime}$,  $\chi_2$ and $\chi^{\prime}_2$ develop nonzero VEV according to, 
\begin{eqnarray}
 \langle\eta_1\rangle=\frac{v_{\eta_1}}{\sqrt2},\,\,
\langle\eta^{\prime}_1\rangle=\frac{v_{\eta^{\prime}_1}}{\sqrt2},\,\,
 \langle \rho \rangle=\frac{v_{\rho}}{\sqrt2},\,\,
  \langle \rho^{\prime} \rangle=\frac{v_{\rho^{\prime}}}{\sqrt2},\,\,\langle\chi_2\rangle=\frac{v_{\chi_2}}{\sqrt2},\,\,
\langle\chi^{\prime}_2\rangle=\frac{v_{\chi^{\prime}_2}}{\sqrt2}.
\label{vevs}
\end{eqnarray}
These VEVs lead to the following gauge symmetry breaking pattern,
\begin{equation}
SU(3)_C \otimes SU(3)_L \otimes U(1)_X  \stackrel{v_{\chi_2}, v_{\chi^{\prime}_2}}{ \Longrightarrow } SU(3)_C \otimes SU(2)_L \otimes U(1)_Y \stackrel{v_{\eta_1}, v_{\eta^{\prime}_1},v_{\rho},v_{\rho^{\prime}} }{ \Longrightarrow } SU(3)_C \otimes U(1)_{QED}.
\label{symmetrygreakingpattern}
\end{equation}

With the breaking of the gauge symmetry by this set of VEVs the expected particles, including the supersymmetric ones, receive mass.  What matter for us here are the scalars' and gauge bosons' masses. Concerning the gauge bosons, they are composed by the standard gauge bosons, $\gamma$, $ Z^0$ and $W^{\pm}$,  two new neutral massive  gauge bosons $Z^{\prime }$ and $U^0$, and two simply charged gauge bosons $V^{\pm}$ with the following mass expression, 
\begin{eqnarray}
\label{gbmass}
 M_{Z^0}^2 &=& \frac{g^2}{4}\frac{(3+4t^2)}{(3+t^2)}(v_\rho^2+v^2_{ \rho^{\prime}}+v_{\eta_1}^2+v_{\eta^{\prime}_1}^{2}), \\
 M_{Z^{\prime }}^2 &=& \frac{g^2}{9}(3+t^2)(v_{\chi_2}^2+v_{\chi_2^{\prime}}^2), \\
 M_{U^0} &=& \frac{g^2}{4}(v_\rho^2+v^2_{ \rho^{\prime}}+v_{\chi_2}^2+v_{\chi_2^{\prime}}^2), \\
 M_{W^\pm} &=& \frac{g^2}{4}(v_\rho^2+v^2_{ \rho^{\prime}}+v_{\eta_1}^2+v_{\eta^{\prime}_1}^{ 2}), \\
 M_{V^\pm} &=& \frac{g^2}{4}(v_{\eta_1}^2+v_{\eta^{\prime}_1}^{ 2}+v_{\chi_2}^2+v_{\chi_2^{\prime}}^2).
\end{eqnarray}
where  $t=g_N/g$,  $v_\rho^2+v^2_{ \rho^{\prime}}+v_{\eta_1}^2+v_{\eta^{\prime}_1}^{ 2}=v^2_{ew}$ and $v_{\chi_2}^2+v_{\chi_2^{\prime}}^2\equiv v^2_\chi$ with $v_\chi$ lying in the TeV scale.

On imposing  the standard relation,
\begin{equation}
 \frac{M_{Z^0}^2}{M_{W^\pm}^2}=\frac{(3+4t^2)}{(3+t^2)}=\frac{1}{\cos^2{\theta_W}},
\end{equation}
we obtain,
\begin{equation}
 t^2 = \frac{\sin^2{\theta_W}}{1-4/3\sin^2{\theta_W}},
\end{equation}
where $\theta_W$ is the electroweak mixing angle. In addition, the mixing between the neutral gauge bosons is given by\footnote{Provided that the mixing among $Z^0$ and $Z^{\prime}$ is very small\cite{Hoang:1996gi}, we neglect such mixing throughout this work.  },
\begin{eqnarray}
 W_N &=& \frac{\sqrt{3}}{\sqrt{3+4t^2}}\gamma-\frac{3t}{\sqrt{3+4t^2}\sqrt{3+t^2}}Z^0+\frac{t}{\sqrt{3+t^2}}Z^{\prime }, \\
 W_8 &=& -\frac{t}{\sqrt{3+4t^2}}\gamma+\frac{\sqrt{3}t^2}{\sqrt{3+4t^2}\sqrt{3+t^2}}Z^0+\frac{\sqrt{3}}{\sqrt{3+t^2}}Z^{\prime }, \\
 W_3 &=& \frac{\sqrt{3}t}{\sqrt{3+4t^2}}\gamma+\frac{\sqrt{3+t^2}}{\sqrt{3+4t^2}}Z^0.
\end{eqnarray}

In order to work in the minimal scenario, we assume R-parity conservation  and invariance  by a $Z_2$ symmetry with the following superfields transforming nontrivially under $Z_2$:  $\left(\hat{l}^c,\hat{d}^c,\hat{u}^c,\hat{\rho},\hat{\rho}^{\prime},\hat{\eta},\hat{\eta}^{\prime} \right) \rightarrow - \left(\hat{l}^c,\hat{d}^c,\hat{u}^c,\hat{\rho},\hat{\rho}^{\prime},\hat{\eta},\hat{\eta}^{\prime} \right)  $. This set of symmetries allows us to work with a shortened superpotential that is formed by the following terms,
 \begin{eqnarray}
  W_{331}  &&= \lambda_{ij}^l   \hat{L}_i \hat{\rho}^{\prime} \hat{l}_j^c
 +  \lambda_{\alpha i}^d   \hat{Q}_{\alpha}\, \hat{\eta}\, \hat{d}_{i_L}^c+
 \lambda_{3i}^d   \hat{Q}_{3}\, \hat{\rho}^{\prime}\, \hat{d}_{i_L}^c+
  \lambda_{\alpha_i}^u   \hat{Q}_{\alpha}\, \hat{\rho}\, \hat{u}_{i_L}^c+   \lambda_{3i}^u   \hat{Q}_{3}\, \hat{\eta}^{\prime}\, \hat{u}_{i_L}^c \nonumber \\
  &&+  \lambda_{\alpha\beta}^{\prime}  \hat{Q}_{\alpha}\, \hat{\chi}\, \hat{d}^{\prime c}_{\beta_L}+
  \lambda_{33}^{\prime}   \hat{Q}_{3}\, \hat{\chi}^{\prime}\, \hat{u}^{\prime c}_{3_L}
  +  f_1\, \varepsilon_{ijk}\,\hat{\eta}^{\prime}_i\, \hat{\rho}^{\prime}_j\hat{\chi}^{\prime}_k    +  f_2\, \varepsilon_{ijk}\,\hat{\eta}_i\, \hat{\rho}_j\hat{\chi}_k \nonumber \\
  &&+  \mu_{\eta}  \hat{\eta} \hat{\eta}^{\prime}   +\mu_{\rho}  \hat{\rho} \hat{\rho}^{\prime}   +  \mu_{\chi}  \hat{\chi} \hat{\chi}^{\prime}+
  \, \mbox{h.c.} \,,
\label{eq:lag-scalar}
\end{eqnarray}
where $\alpha, \beta=1,2$ and $i,j,k=1,2,3$.

Until this point the masses of the ordinary particles are equal to the masses of their superpartners. 
As usual in phenomenological supersymmetric models, SUSY must be broken so as to provide a reasonable shift between ordinary particles and their supersymmetric partners. In this work we assume that SUSY is broken explicitly through the following set of  soft breaking terms that are invariant under the symmetries  assumed here,
\begin{eqnarray}
  {\cal L}_{{Soft}} &&=-\frac{1}{2}\left[ m_{\lambda_G} \sum_{b=1}^8 \left({\bar \lambda}_G^b
 \lambda_G^b  \right)+  m_{\lambda_W} \sum_{b=1}^8 \left( {\bar\lambda}_W^b \lambda_W^b  \right) +  m_{\lambda_X} {\bar\lambda}_X \lambda_X+ h.c.   \right] \nonumber \\
&& + m_{L}^2 \tilde{L}^{\dagger}\tilde{L} + m_{l}^2 \tilde{l}_i^{\dagger}\tilde{l}_i +m_{Q_3}^2 \tilde{Q_3}^{\dagger}\tilde{Q_3}+ m_{Q_{\alpha}}^2 \tilde{Q_{\alpha}}^{\dagger}\tilde{Q_{\alpha}}+ m_{u_i}^2 \tilde{u_i}^{\dagger}\tilde{u_i} \nonumber \\
&&+ m_{u_i}^2 \tilde{u_i}^{\dagger}\tilde{u_i}
 + m_{d_i}^2 \tilde{d_i}^{\dagger}\tilde{d_i}+ m_{u^{\prime}}^2 \tilde{u^{\prime}}^{\dagger}\tilde{u^{\prime}}+ m_{d^{\prime}_{\alpha}}^2 \tilde{d^{\prime}_{\alpha}}^{\dagger}\tilde{d^{\prime}_{\alpha}} 
 - m_{\eta}^2 \eta^{\dagger} \eta - m_{\rho}^2 \rho^{\dagger} \rho \nonumber \\ 
&&- m_{\chi}^2 \chi^{\dagger} \chi  - m_{\eta^{\prime}}^2 \eta^{\prime \dagger} \eta^{\prime} - m_{\rho^{\prime}}^2 \rho^{\prime \dagger} \rho^{\prime} - m_{\chi^{\prime}}^2 \chi^{\prime \dagger} \chi^{\prime}  \nonumber \\
&&+ y^l_{ij}\tilde{L}_i \rho \tilde{l}_{j_L}^c + y^d_{\alpha i} \tilde{Q}_{\alpha} \eta \tilde{d}_{i_L}^c + y^d_{3 i} \tilde{Q}_{3} \rho^{\prime} \tilde{d}_{i_L}^c+ y^u_{\alpha i} \tilde{Q}_{\alpha} \rho \tilde{u}_{i_L}^c \nonumber \\
&&+ y^d_{\alpha i} \tilde{Q}_{\alpha} \eta \tilde{d}_{i_L}^c + y^u_{3i} \tilde{Q}_{3} \eta^{\prime} \tilde{u}_{i_L}^c + y^{\prime}_{\alpha i} \tilde{Q}_{\alpha} \chi \tilde{d}^{\prime c}_{i_L}+ y^u_{33} \tilde{Q}_{3} \chi^{\prime} {\tilde{u}^{\prime c}}_{3_L}\nonumber \\
&&- \left[ k_1 \varepsilon_{ijk} \eta_i \rho_j \chi_k +k_2 \varepsilon_{ijk} \eta^{\prime}_i \rho^{\prime}_j \chi^{\prime}_k + h. c.  \right] \nonumber \\
&&+ b_{\eta}\, \eta^{\prime}\eta + b_{\chi}\, \chi^{\prime} \chi + b_{\rho}\, \rho^{\prime}\rho\,,
\label{softterms}
 \end{eqnarray}
where $\lambda_G^b$ are the gluinos, $\lambda_W^b$ are gauginos associated to $SU(3)_L$ (in both cases $b$ is the gauge group index) and $\lambda_X$ is the gaugino associated to $U(1)_X$,  scalar supersymmetric partners of fermion fields, $f$, are denoted by $\tilde{f}$, while the remaining fields are self-evident.

Once we have settled the interactions and parameters of the S331RH$\nu$ model, we are, then, ready to start the development  of the main proposal of this work  that is to check if the R-sneutrino of the model is a good CDM. But first  we study the possible range of parameters that can explain the observed 125~GeV Higgs mass. This will constrain the parameters that will be used in the CDM analysis.

\section{ Higgs mass: numerical results}
\label{sec2}

In this section we  obtain  the mass of the lightest CP-even neutral scalar provided by the model, which we assume is the Higgs boson. First of all we have to obtain the scalar potential which  is composed by $V=V_F + V_D + V_{soft}$ where $V_F$ and $V_D$ are  the F-term and D-term, respectively, and $V_{soft}$  comes from the soft SUSY breaking terms.  
With the potential in hand we are ready  to obtain the minimum conditions over the potential which translate to a  set of constraint equations, $\frac{\partial V}{\partial \phi_i}|_{\phi_i = \langle \phi_i \rangle_0}=0$, where ${\phi_i = \langle \phi_i \rangle_0}$ means that all scalar fields are computed at their VEV. The squared mass matrix can then be built by taking $\frac{\partial^2 V}{\partial \phi_a \partial \phi_b}|_{\phi_i = \langle \phi_i \rangle_0}$.
Finally, by applying the set of minimum conditions over the mass matrices and diagonalizing them  we obtain the physical scalars of the model. Due to the complexity of $V$ we are not showing the analytical expressions for   $\frac{\partial V}{\partial \phi_i}|_{\phi_i = \langle \phi_i \rangle_0}=0$ and $\frac{\partial^2V}{\partial \phi_a \partial \phi_b}|_{\phi_i = \langle \phi_i \rangle_0}$ here, which are not illuminating at all. We then proceed with a numerical approach to diagonalize the mass matrices in question.  For this we made  use of a subroutine  called jacobi \cite{Press:2007:NRE:1403886} which composes the micrOMEGAs  package~\cite{Belanger:2008sj,micro}. This is enough to develop the features of the model we are interested in.

The CP-even neutral scalar fields compose  a $10\times10$ mass matrix. However, the neutral scalars $\eta_2\,,\,\eta_2^{\prime}\,,\, \chi_1\,,\,\chi^{\prime}_1$ carry two units of lepton number and, as far as lepton number is conserved, they decouple from the other six neutral scalars. On diagonalizing the remaining $6 \times 6$ mass matrix we obtain six physical CP-even neutral scalars.  Two of them, which is  combination mainly  of $\chi_2$ and $\chi^{\prime}_2$,  are very heavy with mass at 3-3-1 scale, typically around few TeV. The other four, which are mainly  combinations of $\eta_1$ $\eta_1^{\prime}$, $\rho$ and $\rho^{\prime}$,  acquire masses at electroweak scale with the lightest of them being the Higgs. We refer to these  four  scalars as $h$ (the Higgs boson), $h^{\prime}$, $H$ and $H^{\prime}$.

In what follows we present the results only for the lightest CP-even scalar, the Higgs boson. For this  we choose as independent parameters  the following set of variables, where their range of values to be scanned in the numerical computation were fixed so as to guarantee the scalar potential stability, 
\begin{eqnarray}
\centering
0.0001 &\leq & |f_1,f_2| \leq 0.0049\,,\,\,\,\,\,\,\,\,\,\,8\,\mathrm{GeV} \leq |k_1 ,k_2| \leq 15\mathrm{GeV}\,,\nonumber \\
400\,\mathrm{GeV} & \leq & |\mu_{\eta},\mu_{\rho} | \leq 700\,\mathrm{GeV}\,,\,\,\,\,\,\,\,\,\,\,
800\,\mathrm{GeV} \leq |\mu_{\chi}|\leq 10000\,\mathrm{GeV}\,, 
\nonumber \\
300\,\mathrm{GeV}^2\ & \leq & |b_{\eta}, b_{\rho} | \leq 500\,\mathrm{GeV}^2\,,\,\,\,\,\,\,\,\,\,\,
50000\,\mathrm{GeV}^2\leq | b_{\chi}|\leq 100000\,\mathrm{GeV}^2 \,,
\nonumber \\
40 &\leq & v_{\eta_1}\leq 140\,\mathrm{GeV}\,,\,\,\,\,\,\,\,\,\,\, 30\,\mathrm{GeV}\leq v_{\eta^{\prime}_1},v_{\rho^{\prime}}\leq 50\,\mathrm{GeV}\,,
\nonumber \\
5000\,\mathrm{GeV} &\leq &   v_{\chi_2}\leq 10000\,\mathrm{GeV}\,,\,\,\,\,\,\,\,\,\,\,700\,\mathrm{GeV}\leq   v_{\chi^{\prime}_2}\leq 2000\,\mathrm{GeV}\,,
\end{eqnarray}
while there is a constraint among some of the VEVs,
\[
v_{\rho}^2+v_{\eta_1}^2+v_{\eta\prime_1}^2+v_{\rho\prime}^2=(246\,\mathrm{GeV})^2\,,
\]
which comes from the known $W^{\pm}$ mass.

From the numerical diagonalization of the $6 \times 6$ mass matrix we obtain that the lightest CP-even neutral scalar gains mass at tree level in the range from $80$~GeV to $100$~GeV. Considering that in the MSSM the maximal value the Higgs mass may attain at tree level is $91$ GeV, we have that  the S331RH$\nu$ model provides a better tree level contribution to the Higgs mass. However, loop corrections to the Higgs mass are still necessary. At this point we just consider the leading one-loop correction for the Higgs mass dictated by the MSSM whose expression is~\footnote{While the S331RH$\nu$ model contains the MSSM, justifying this approximation, we remark that a finer computation can be pursued not only by including full two-loop effects~\cite{Hahn:2013ria} but also the new contributions specific from the enlarged particle spectrum of the model.}, 
\begin{equation}
\Delta m^{2}_{h}=
\frac{3m^{4}_{t}}{2\pi^{2}v_{ew}^{2}}\left(log\left(\frac{M^{2}_{s}}{m^{2}_{t}}\right)
+
\frac{X^{2}_{t}}{M^{2}_{s}}\left(1-\frac{X^{2}_{t}}{12M^{2}_{s}}\right)\right),
\label{Limh2}
\end{equation}
where $m_{t}$ is the top mass, $v_{ew}=246$~GeV is the standard electroweak VEV,  $X_{t}$ is the soft trilinear coupling of the stops and $M_{s} \equiv
(m_{\tilde{t}_{1}}m_{\tilde{t}_{2}})^{1/2}$ is the SUSY scale (scale of
superpartners masses) where $m_{\tilde t}$ is the stops' mass, that we suppose to be degenerated.

\begin{figure}[!h]
\centering
\includegraphics[scale=.50]{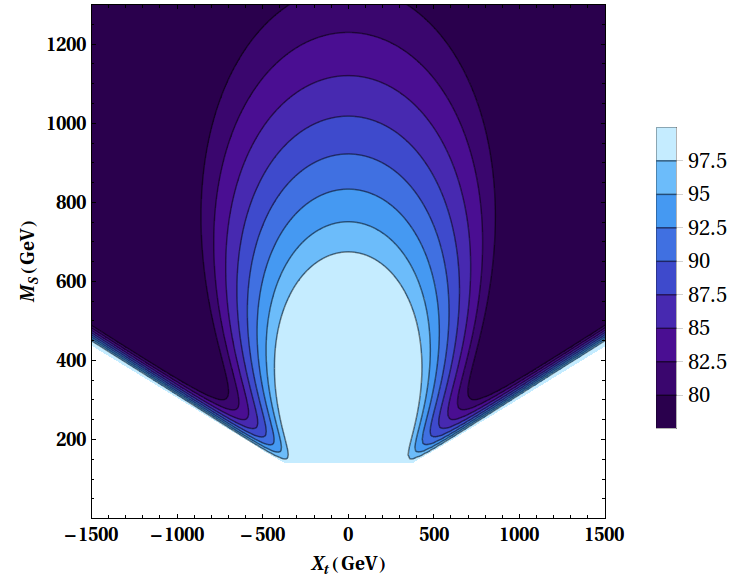}
\caption{Contour plot  corresponding to $m_h=125$GeV in the $M_s \,\mbox {versus} \, X_t $ plane  where  $m_t=173$ GeV and $v_{ew}=246$ GeV. The legend bar indicates the range of values provided solely by the  tree level mass.}
\label{figloopHiggs}
\end{figure}

We add this one-loop contribution to the tree level Higgs mass and then perform the scan on the parameter space. Our  results are  shown in  Fig.~(\ref{figloopHiggs}). 

It is remarkable that the S331RH$\nu$ model is able to yield a tree level Higgs mass around $100$ GeV (the lightest blue in Fig.~(\ref{figloopHiggs})), where stop mass, $m_{\tilde{t}}$, below TeV along with a small $X_t$  are enough  to generate the necessary radiative corrections to produce the observed Higgs mass at one loop in the approximation where only stops were taken into account.
In other words, differently from MSSM where $m_{\tilde t}$ is pushed beyond 1~TeV and $X_t$ is rather large, there is no tight constraints on these parameters in the S331RH$\nu$ model, which can easily  accommodate a 125~GeV Higgs mass. 

This result is not a surprise at all. Extensions of the MSSM that present cubic invariant terms in the superpotential generate new contributions to the Higgs potential of the MSSM which, consequently, result in new corrections at tree level  to  the Higgs mass. For example,  in the NMSSM  the superfield singlet $\hat \phi$ is added to the MSSM superfield content and composes with the standard superfields $\hat H_u$ and $\hat H_d$  the following  invariant cubic term  $\lambda \hat H_u \hat H_d \hat \phi$ in the superpotential of the model. Such term furnishes an additional  tree level correction to the  Higgs mass expression of the MSSM which is possible to lift the Higgs mass by some units of GeV which is sufficient to alleviate the tension on the quantum corrections involving stops \cite{Hall:2011aa}. Another  example is the extension of the MSSM with the superfileds triplets  $\hat \Delta_1$ and $\hat \Delta_2$. In this case the cubic invariant terms
 $\lambda_1\hat H_u \hat \Delta_1 \hat H_u + \lambda_2 \hat H_d \hat \Delta_2 \hat H_d $ compose the superpotential of the model and provide robust   tree level corrections to the Higgs mass\cite{FileviezPerez:2012ab}. In the particular  case of 3-3-1  models, the Higgs sector usually involves  three Higgs triplets. When this is the case, cubic invariant terms as  $f_1$ and $f_2$  given in  Eq. (\ref{eq:lag-scalar}) compose  the superpotential of the supersymmetric versions of these models.  Consequently these terms  will generate new corrections  at tree level to the Higgs mass  predicted by the MSSM. This was firstly perceived in \cite{Duong:1993zn}. Our numerical approach here is in agreement with such predictions.

Guaranteeing that our model recovers the observed Higgs boson mass, in the next section we use the same set of parameters scan to examine if $\tilde \nu_R$ is viable as CDM candidate.

\section{Relic Abundance and Direct Detection}
\label{sec3}
In a SUSY model where R-parity is conserved, the lightest supersymmetric particle (LSP) is the natural candidate for CDM~\cite{Jungman:1995df,Bertone:2004pz}. In the MSSM, the CDM may be a scalar, the superpartner of the  left-handed neutrino,   $\tilde \nu_L$, or a combination of Majorana fermionic superpartners of the scalars and the Z boson, the neutralinos. However, we already pointed out the reason why $\tilde \nu_L$ is not a viable CDM candidate, and then the MSSM inevitably offers only neutralinos to play this role.

Nevertheless, extensions or variants of MSSM do allow sneutrinos as CDM, which happens when right handed neutrinos are somehow part of the field content to be supersymmetrized~\cite{Asaka:2005cn,Gopalakrishna:2006kr,Asaka:2006fs,
Lee:2007mt,Arina:2007tm,Cerdeno:2008ep,Cerdeno:2009dv,Belanger:2010cd,Belanger:2011rs,Dumont:2012ee,
DeRomeri:2012qd,Tang:2016oba}. In this case, generally, a mixing among right handed and left handed sneutrino may be the LSP and constitute the CDM candidate. 

In the S331RH$\nu$ model, in addition to $\tilde \nu_L$ and neutralinos, we have a third possibility in the form of scalar right handed neutrino (or simply R-sneutrino), $\tilde \nu_R$, which emerges naturally in this model as the third component in the leptonic triplet of $SU(3)_L$. Since the $SU(2)_L$ subgroup of $SU(3)_L$ contains the  matter content of MSSM, the same conclusions over the CDM candidates derived there apply to the $\tilde{\nu}_L$ in our model. We are then left with neutralinos or R-sneutrino, which possibly can play the CDM role. 
Both were already investigated in a similar SUSY model with different scalar content and assumptions in Ref.~\cite{Long:2007ev,Huong:2008ww}. Besides, their analysis on the R-sneutrino was taken considering it as self-interacting dark matter and cannot be compared with ours.
There is a reasonable complication in our neutralino spectrum compared to the MSSM, which is related to the larger higgsino as well as gaugino spectrum of S331RH$\nu$ model, as can be seen from Eqs.~(\ref{part_transf1}),~(\ref{part_transf2}),~(\ref{gbmass}). The resulting neutralino mass eigenstate amounts to diagonalize a $15 \times 15$ mass matrix in contrast to the $4 \times 4$ mass matrix of MSSM. In this work we are not considering the neutralinos as CDM though. Instead, our interest is driven to the LSP as the R-sneutrino. 

In what concerns $\tilde \nu_R$, we stress that  in the S331RH$\nu$ model as well as in the MSSM neutrinos gain mass through effective operators. The gauge and discrete, $Z_2$, symmetries assumed in this work allow for the following effective operators as  source of neutrino masses,
\begin{equation}
\frac{\lambda^{\nu_L}}{\Lambda}(\hat L \hat \eta ^{\prime})(\hat \eta^{\prime} \hat L) + \frac{\lambda^{\nu_R}}{\Lambda}(\hat L \hat \chi ^{\prime})(\hat \chi^{\prime} \hat L)\,,
\label{effecticeoperators}
\end{equation}
where $\lambda^{\nu_L}$ and $\lambda^{\nu_R}$ are dimensionless parameters and $\Lambda$ is a grand unification mass scale~\footnote{In the numerical computations we will take the R-sneutrino mass as a free parameter, varying other parameters like soft masses and VEVs that constrain the former ones in order to obtain the correct active neutrinos'  masses.}.
Notice that the first effective operator engenders a mass term to the $\nu_L$, since only the scalar component, $\eta^\prime_1$ of $\hat{\eta}^\prime$ develops VEV, while the second operator gives mass to $\nu_R$, in this case because only the $\chi^\prime_2$ scalar component of $\hat{\chi}^\prime$ develops VEV.
This implies that the left-handed neutrinos do not mix with the right-handed ones. Also, they are completely sterile in relation to the standard gauge boson interactions as they interact solely with the gauge bosons of the 3-3-1 symmetry, namely, $V^{\pm}$, $Z^{\prime }$ and $U^0$. These properties are inherited by R-sneutrinos and, consequently, $\tilde \nu_R$ does not mix with $\tilde \nu_L$ too. Besides, the bridge between them and SM particles is made through $Z^\prime$ and the scalars.  All these features make $\tilde \nu_R$ rather distinct from the usual MSSM extensions where R-sneutrino is the CDM candidate, justifying and further motivating our analysis of $\tilde \nu_R$ in this context.  Finally, it is important to say that,  as far as we know,  this is the first time that $\tilde \nu_R$ is considered as a WIMP in the framework of S331RH$\nu$ model. We compute its relic abundance and direct detection constraints in the following subsections.

\subsection{Relic Abundance}

It is well known that the  relic abundance of a WIMP  is directly related to its thermal averaged annihilation cross section at the time of freeze-out~\cite{Jungman:1995df,Bertone:2004pz}. Its  decoupling is roughly determined when the interaction rate drops below the expansion rate of the universe. In order to obtain the WIMP's abundance we have to solve the Boltzmann equation,   
\begin{equation}
\frac{dY}{dT}=\sqrt{\frac{\pi g_{\ast}(T)}{45}}M_{p} -\left\langle \sigma v\right\rangle(Y^{2}-Y^{2}_{eq})\,,
\label{relicabundance}
\end{equation}
which gives the evolution of the abundance of a generic species in the universe. In it  $Y$ is the relic abundance as function of the temperature, $T$, of the thermal bath,  $Y_{eq}$ the thermal equilibrium abundance, $g_{\ast}$ is the effective number of degrees of freedom at thermal equilibrium, $M_{p}$ is the Plank mass.  $\left\langle \sigma v\right\rangle$ is the thermal averaged cross section for WIMP annihilation, with $v$  the relative velocity between the annihilating particles. It is in this cross section that the particle physics modeling gets into the scene, and its expression can be written as~\cite{Belanger:2008sj,micro},
\begin{equation}       
\left\langle \sigma v\right\rangle=\frac{\displaystyle\sum_{i,j}g_{i}g_{j}\displaystyle\int_{ (m_{i}+m_{j})^{2} }ds\sqrt{s}K_{1}(\frac{\sqrt{s}}{T})p_{ij}^{2}\displaystyle\sum_{k,l}\sigma_{ij;kl}(s)}{2T(\displaystyle\sum_{i}g_{i}m^{2}_{i}K_{2}(m_{i}/T) )^{2}}\,,
\label{evolutionabundance}
\end{equation}
where $g_{i}$ is the number of degrees of freedom of the species involved, $\sigma_{ij;kl}$ the total cross-section for annihilation of a pair of particles with masses $m_{i}$, $m_{j}$ into some SM particles $(k,l)$ of masses $m_{k}$, $m_{l}$, $p_{ij}$ is the momentum of incoming particles in their center of mass frame, with squared total energy, $s$, and the functions $K_1$ and $K_2$ are modified Bessel functions of first and second kind, respectively. 

The relic density is obtained integrating from $T=\infty$ to $T=T_{0}$, where $T_{0}$ is the temperature of the Universe today, precisely measured by the cosmic microwave backgraound radiation (CMBR) spectrum~\cite{Hinshaw:2012aka,Ade:2015xua}. It can be cast as~\cite{Belanger:2008sj,micro},
\begin{equation} 
\Omega h^{2}=2.742\times 10^{8}\frac{M_{WIMP}}{\mbox{GeV}} Y(T_{0})\,.
\label{solution}
\end{equation}

Given the large amount of interactions and mass diagonalization required in the model, an analytical approach to compute the relic abundance is unfeasible. Instead, we opt for a numerical computation using  the codes: LanHEP~\cite{Semenov:2014rea} to generate the Feynman rules in an CalcHEP~\cite{calchep} output to be called in micrOMEGAs~\cite{Belanger:2008sj,micro}. The micrOMEGAs code is very useful in computing the CDM abundance, including coannihilation. In addition, by means of CalcHEP, it allows us to calculate the CDM scattering cross section normalized to nucleon, so we can compare with exclusion plots given by recent direct detection experiments~\cite{Agnese:2013rvf,Aprile:2012nq,Reindl:2015dnp,Akerib:2015rjg}. 

\begin{figure}[!h]
\centering
\includegraphics[scale=0.45]{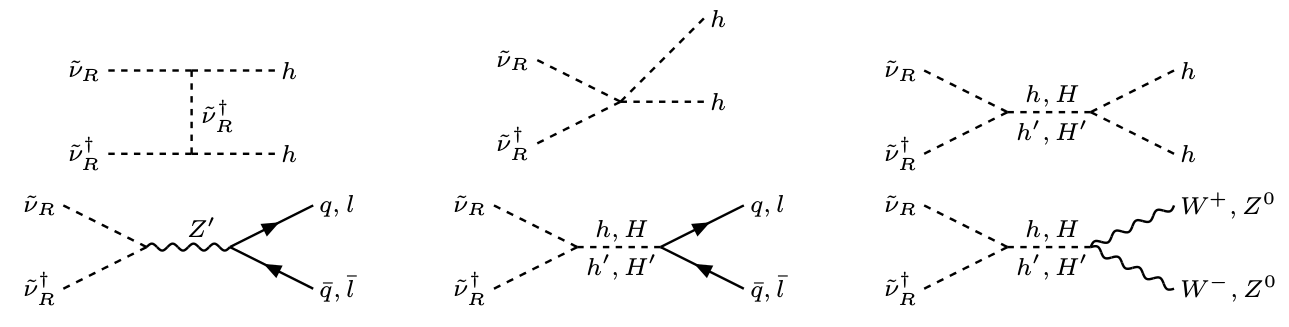}
\caption{Dominant processes contributing to the R-sneutrino abundance. $q$ and $l$ means quarks and leptons, respectively.}
\label{diagomg}
\end{figure}

From now on, for simplicity, we will assume that right-handed neutrinos and sneutrinos are in a diagonal basis, 
and will consider that the lightest of the R-sneutrinos is our WIMP. 
We start by presenting the main channels involved in the CDM annihilation cross section, where the relevant interactions are mediated by Higgs and $Z^\prime$, as can be seen in Fig.~(\ref{diagomg}).

\begin{figure}[ht]
\centering
 \includegraphics[width=0.49\textwidth]{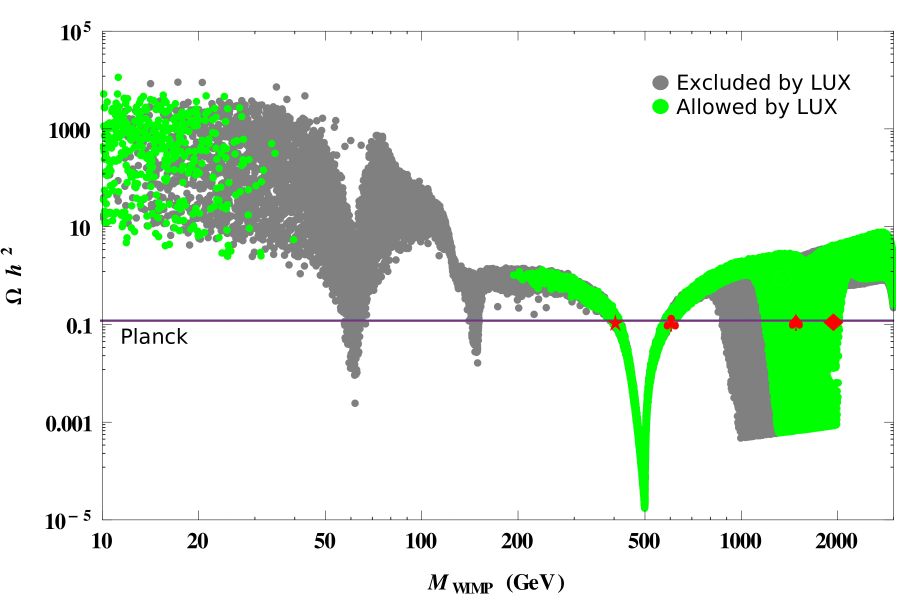}
  \includegraphics[width=0.49\textwidth]{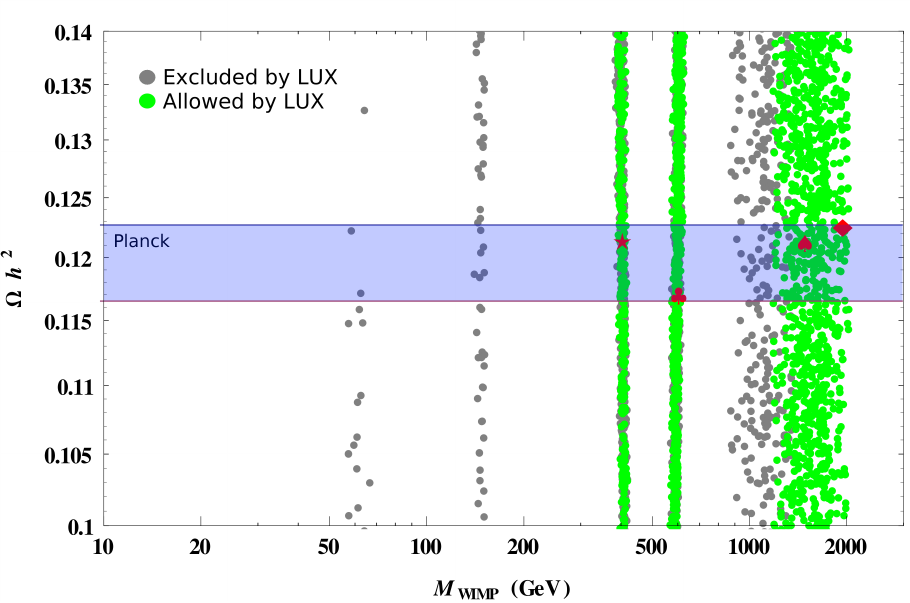}
 \caption{Relic density versus WIMP mass. Right panel is  an improvement in resolution around the Planck bounds. The gray dots are ruled out by direct detection data. The green dots are in accordance with LUX bounds~\cite{Akerib:2015rjg}, while the region enclosed by purple lines represents the Planck constraints (blue shaded region on the right panel)~\cite{Ade:2015xua}.}
 \label{rot}
\end{figure}

In the Fig. (\ref{rot}) we show the results of the R-sneutrino relic density.  Observing the dips in the scatter plot presented in the left panel, one clearly recognizes the resonances,  $h$, $H$, $h^\prime$, $H^\prime$ and $Z^\prime$, whose masses are $m_{h}=125$~GeV, $m_{H}=300$~GeV, $m_{h^\prime}=m_{H^\prime}=1000$~GeV and $2000$~GeV $< m_{Z^\prime} < 4000$~GeV, respectively (these phenomenological reasonable values for the heavier scalars were fixed for simplicity, although they could also be varied). In the right panel, we show the same results as the left panel but zoomed in the region in the vicinity of the relic density as observed by the Planck satellite~\cite{Ade:2015xua}. In these plots, we have included the direct detection results provided by LUX~\cite{Akerib:2015rjg}, which are going to be better explained in the next section. The gray region is excluded and the green region is allowed by LUX results. In addition, in order to provide the precise values of the parameters involved in the process, and the dominant channels in different DM mass regions, four benchmark points were included in all plots, given by the table~\ref{bench}, all in agreement with the constraints mentioned before.

In principle, we have four possible regions providing the correct abundance, where preferred WIMP mass is about 60~GeV~\footnote{Such a low mass would be ruled out by the preference of Higgs to decay into two WIMPs in this model, as could be directly inferred from the investigation in Ref.~\cite{Alvares:2012qv}.}, 150~GeV, 500~GeV and between 1000~GeV and 2000~GeV. However, when we take into account the  direct detection bounds we restrict this region just to about 500~GeV (scalar resonances) and between 1000 and 2000 GeV ($Z^\prime$ resonance). 

In the next section, we will detail the CDM scattering cross section to nucleon and show the results obtained for our model including the complementary CDM relic density.
\begin{table}[!h]
{\bf Benchmark Points}\\ 
\vspace{0.3cm}
\begin{tabular}{|c|c|c|c|c|c|}
\hline
Symbol & $M_{ \tilde{\nu}_R}$ (GeV) & $\Omega h^2$ & $\sigma_{ \tilde{\nu}_R-n}$ (pb) & $M_{Z^{ \prime }}$ (GeV) & Main Channels\\
\hline
{\color{red} \ding{72}} & 400 &  $\ \ 0.1215 \ \ $ & $4.041\times10^{-9}$ & 3630 & $\tilde{\nu}_R, \tilde{\nu}_R \stackrel{h^\prime, H^\prime}{ \longrightarrow }  W^\pm, Z^{ 0 },Z^{ 0 }$\\
\hline
{\color{red} \ding{168}} & 600 & 0.1171 & $4.249\times10^{-9}$ & 3597 & $\tilde{\nu}_R, \tilde{\nu}_R \stackrel{h^\prime, H^\prime}{ \longrightarrow }  W^\pm, Z^{ 0 },Z^{ 0 }$\\
\hline
{\color{red} \ding{171}} & 1480 & 0.1213 &  $5.463\times10^{-9}$ & 3400 & $ \ \ \tilde{\nu}_R, \tilde{\nu}_R \stackrel{Z^\prime}{ \longrightarrow } \bar{q}, q \ \ $ \\
\hline
{\color{red} \ding{169}} & 1934 & 0.1226 & $3.400\times10^{-9}$ & 3819 & $\tilde{\nu}_R, \tilde{\nu}_R \stackrel{Z^\prime}{ \longrightarrow } \bar{q}, q $\\
\hline
\end{tabular}
\caption{In this table we show some points with specific values for parameters of the model and the dominant  channels.}
\label{bench}
\end{table}

\subsection{Direct Detection}

\begin{figure}[!h]
\centering
\includegraphics[scale=0.5]{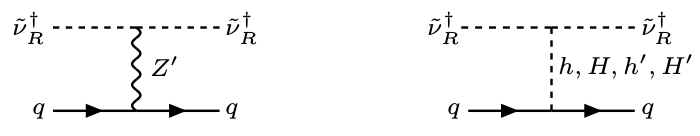}
\caption{Dominant processes to the WIMP-nucleon scattering cross section.}
\label{diag2}
\end{figure}

It is a well motivated hope that, generally, any possible CDM candidate may interact with the target nuclei  (more specifically the nucleons) of the detectors. These interactions may be axial, referred as spin-dependent interactions (SD), scalar and/or vector like, known as spin-independent interactions (SI). In our model, the principal channels providing considerable direct detection rates are given by Higgs particles and $Z^\prime$ (see Fig. (\ref{diag2})), meaning we have just SI interactions. 

The effective lagrangian for SI contributions is given by,
\begin{equation}
 \mathcal{L} \supset \alpha^S_q \, \tilde{\nu}_R^\dagger \tilde{\nu}_R \, \, \bar{q} q + \alpha^V_q \, \tilde{\nu}_R^\dagger \partial_\mu \tilde{\nu}_R \, \, \bar{q} \gamma^\mu q ,
\end{equation}
with the couplings $\alpha^S_q$ and $\alpha^V_q$ depending  on the parameters of the model. The WIMP-nucleus cross section that can be derived from this lagrangian is~\cite{Belanger:2008sj},
\begin{equation}
\sigma_0 = \frac{4\mu_N^2}{\pi}[Z f^p + (A-Z) f^n]^2\,,
\label{SICS}
\end{equation}
where $\mu_N$ is the WIMP-nucleus reduced mass, $Z$ is the number of protons and $A$ is the number of nucleons. The function $f^{p,n}$ is the WIMP-nucleon amplitude that carries the particle physics model information which, for the proton, is given by~\footnote{We restrict ourselves to the scalar interaction since the experimental results are parametrized by this contribution to the WIMP-nucleon cross section.},
\begin{equation}
\frac{f^p}{m_p}=\sum_{q=u,d,s}\frac{\alpha^S_q}{m_q} f^{p}_{Tq}+\frac{2}{27}f^{p}_{TG}\sum_{q=c,b,t}\frac{\alpha^S_q}{m_q}\,,
\label{fp}
\end{equation}
where the coefficients $f^p_{Tq}$ and $f^{p}_{TG}$ are the contributions of light quarks to the proton mass, $m_pf^p_{Tq}=\langle p|m_q \bar{q}q|p\rangle$, and the WIMP-gluon interaction through quark loops, respectively, with $f^{p}_{TG} = 1-\sum_{q=u,d,s}f^{p}_{Tq}$. Experimentally we have,
\begin{equation}
f^p_{Tu}=0.020\pm 0.004\,,\,\,\,\,\,\,\,\,\,\,f^p_{Td}=0.026\pm 0.005\,,\,\,\,\,\,\,\,\,\,\,f^p_{Ts}=0.118\pm 0.062\,.
\label{factors}
\end{equation}
The expression for $f^{n}$ can be easily obtained taking into account that, $f^n_{Tu}=f^p_{Td}$, $f^n_{Td}=f^p_{Tu}$ and $f^n_{Ts}=f^p_{Ts}$. We then can write the Wimp-nucleon scalar cross section that is useful for comparison with experimental results as,
\begin{equation}
 \left(\frac{d\sigma_{Wimp-nucleon}}{dE_R}\right)_{SI} = \frac{m_{N} \sigma_{p,n}}{2 \mu_{p,n}^2 v^2}\frac{[Z f^p + (A-Z) f^n]^2}{(f^{p,n})^2} F^2(E_R)\,,
 \label{cswn}
\end{equation}
where $F^2(E_R)$ is the nuclear form factor, $E_R$ is the nucleus recoil energy, $v$ is the WIMP velocity, $\mu_{p,n}$ is the WIMP-nucleon reduced mass and $\sigma_{p,n}$ is given by,
\begin{equation}
\sigma_{p,n} = \frac{4\mu^2_{p,n}}{\pi}(f^{p,n})^2\,.
\end{equation} 
For detailed steps leading to the cross section in Eq.~(\ref{cswn}) above we indicate Refs.~\cite{Jungman:1995df,Belanger:2008sj,Bertone:1235368}. 
\begin{figure}[!h]
\centering
 \includegraphics[scale=1.2]{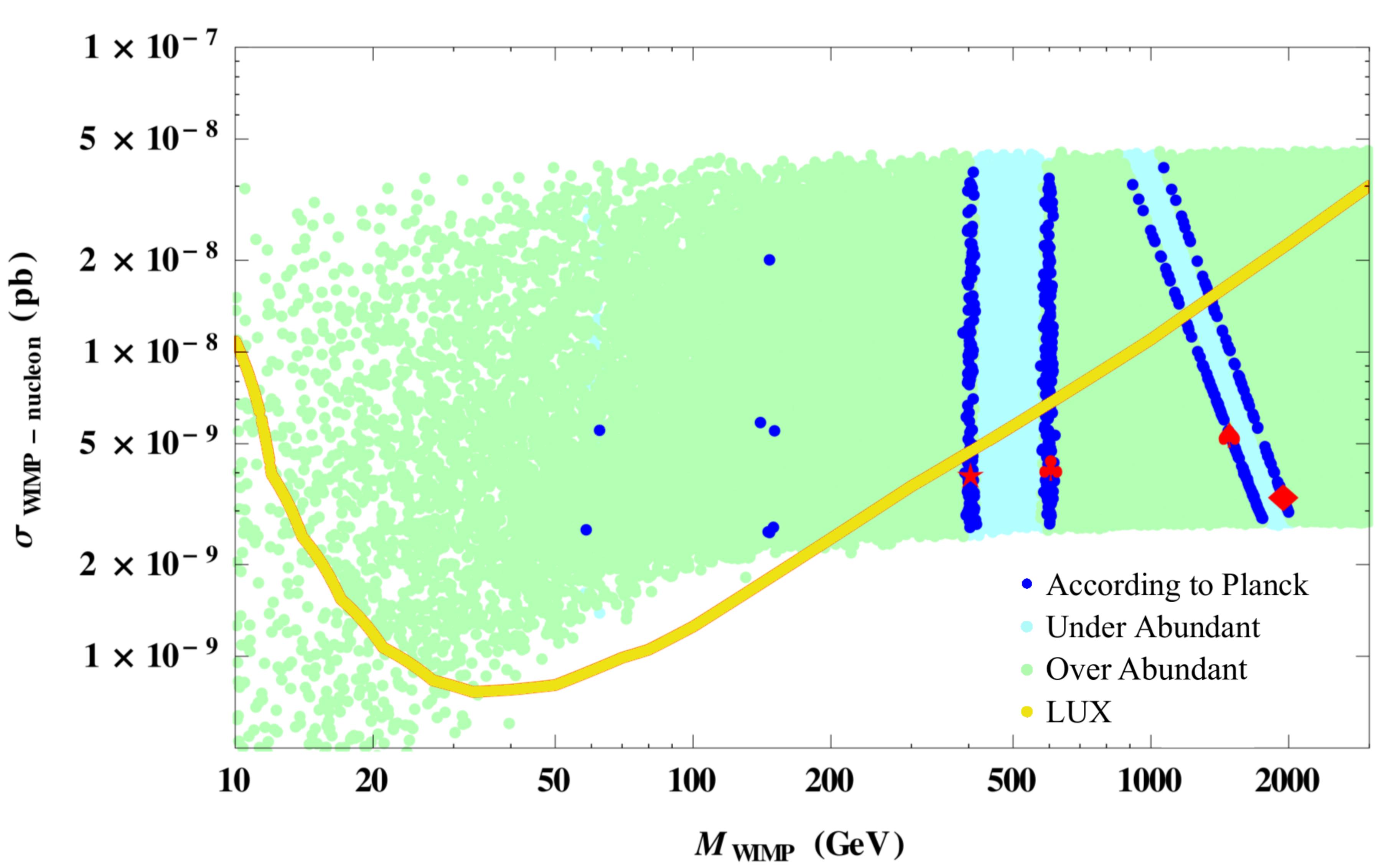}
 \caption{WIMP-nucleon cross section versus dark matter mass. In this plot the green points indicate overabundance, the blue points are those in agreement with the Planck bounds, and light blue points correspond to an abundance lower than needed to explain all CDM. All points above the yellow curve are excluded by direct detection from LUX~\cite{Akerib:2015rjg}. }
 \label{cs}
\end{figure}

Once again, in order to perform the numerical computation and obtain the elastic scattering WIMP-nucleon cross section  for the S331RH$\nu$ model we use the numerical package micrOMEGAS~\cite{Belanger:2008sj,micro}. We present our  results in Fig.~(\ref{cs}) in the plane WIMP-nucleon cross section versus WIMP mass. In this plot, the yellow line represents the upper bound on CDM cross section provided by LUX~\cite{Akerib:2015rjg}, again we use the complementary abundance constraints.  The region in light green is overabundant, light blue is underabundant and blue are in agreement with the cosmological CDM abundance. Here, we included the same Benchmark Points presented in table~\ref{bench}.  Observe that the blue dots follows the resonance regions mentioned before. It is also important to emphasize that the direct detection puts the following lower bound on the $\tilde \nu_R$ mass ($m_{\tilde \nu_R} \geq 400$~GeV).

\begin{figure}[ht]
\centering
 \includegraphics[scale=1.2]{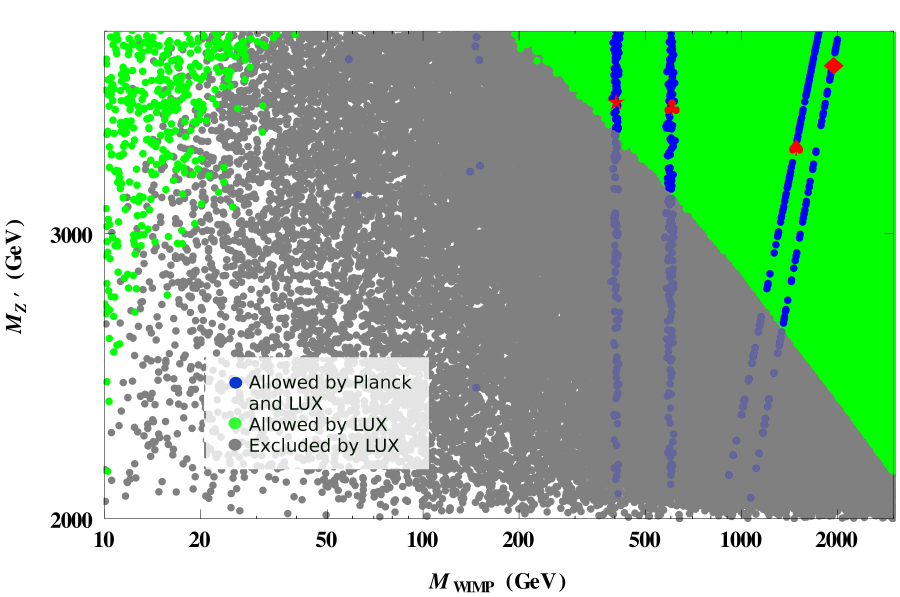}
 \caption{ $Z^{\prime}$ mass versus WIMP mass. The blue points furnish the correct abundance as indicated by Plank~\cite{Ade:2015xua}. The green region is in agreement with recent direct detection experiment LUX~\cite{Akerib:2015rjg} and the gray region is excluded by it.}
 \label{mzp}
\end{figure}

As our last result, we obtain the constraint coming from CDM observables on $Z^\prime$ mass. The results are presented in Fig.~(\ref{mzp}).  The gray points are ruled out by LUX~\cite{Akerib:2015rjg}, while the green points lie in the allowed region of the parameter space. The blue points provide the observed values for CDM relic density from Planck~\cite{Ade:2015xua}. As we can see, the LUX constraints on elastic WIMP-nucleon scattering cross section along with the correct relic density observed by Planck are able to establish a lower bound on $Z^\prime$ mass, $m_{Z^\prime} \gtrsim 2400$~GeV, compatible with a model independent analysis performed in Ref.~\cite{Alves:2015pea}, as well as a particular 331 model with left handed neutrinos in the leptonic triplet~\cite{Profumo:2013sca}.  Besides, this result is close to LHC constraints on a non SUSY 331 model with right handed neutrinos that impose $Z^\prime$ mass to lie above $M_{Z^\prime} \gtrsim 2200$~GeV~\cite{Coutinho:2013lta}, which can be further investigated in the context of the S331RH$\nu$ in future work.

\section{Conclusions}
\label{sec4}
We have built a SUSY version of the gauge $SU(3)_c\otimes SU(3)_L \otimes U(1)_X$ model with right handed neutrinos with three scalar triplets, the S331RH$\nu$ model. Our first aim was to show that a Higgs boson mass of 125~GeV can be obtained without the tight bounds on stop mass and soft trilinear coupling, usually required in MSSM versions. Since our model was able to generate a tree level Higgs mass between 80~GeV to almost 100~GeV, the loop corrections coming from stops were alleviated, demanding stop mass as low as 200~GeV at one loop leading order and never much higher than 1~TeV for extremely low (close to zero) soft trilinear coupling. By itself that is already an appealing motivation to develop this model. 

We also enjoyed the opportunity to investigate the right handed sneutrino, $\tilde{\nu}_R$, as a CDM candidate since the model has it for free in its particle multiplets. The $\tilde{\nu}_R$ in this kind of gauge model was never studied as a WIMP, only scarcely the neutralinos were considered and in a rather different version of this model as a matter of fact~\cite{Huong:2008ww}, although these are a little more intricate here as it involves a mixing of 15 neutral particles. Then, the S331RH$\nu$ model offers two possibilities of WIMPs, but we concentrated on sneutrinos because it is simpler to handle than neutralinos, besides being a natural possibility in this model, not easy to attain in every SUSY model. We have computed its relic abundance, contrasted with Plank observed CDM density, and direct detection bounds from LUX experiment. We have analyzed a large portion of the parameter space, highlighting some benchmark points and our results have shown that $\tilde{\nu}_R$ is a viable WIMP if its mass is above $400$~GeV, which makes it a very interesting WIMP to be searched at LHC.

Finally, since the right handed sneutrino couples to a new neutral gauge bozon, $Z^\prime$, we pushed our CDM search to put some bounds on  $Z^\prime$ mass. Assuming that $\tilde{\nu}_R$ is the only CDM component (or at least the one that corresponds to almost all CDM observed), the Planck results together with LUX exclusion plots allowed us to impose bound on the plane WIMP mass against $Z^\prime$ mass, implying a lower bound $M_{Z^\prime} \gtrsim 2400$~GeV, in consonance with existing bounds on the non-supersymmetric version of this model coming from LHC searches on $Z^\prime$.

All of this constitute interesting outcomes of this supersymmetric model that contains several theoretical features to be further explored, besides being phenomenologically testable at LHC, as well as current experiments on CDM direct and indirect detection, which we intend to explore soon.

\acknowledgments

The authors would like to thank Alexandre Alves for a careful reading of the manuscript and useful suggestions and J. G. Ferreira Jr and Jamerson Rodrigues for useful discussions. This work was supported by Conselho Nacional de Pesquisa e
Desenvolvimento Cient\'{i}fico- CNPq (C.A.S.P,  P.S.R.S. and C.S. ) and Coordena\c c\~ao de Aperfei\c coamento de Pessoal de N\'{i}vel Superior - CAPES (A.C.O.S). 
\appendix
\section{Relevant interactions}
 
The relevant interaction terms  involving R-sneutrino, Higgs and $Z^{\prime}$ that matter for the calculation of abundance and scatter cross sections are given by,
  
 \begin{equation}
  {\cal L} \supset  - \frac{g \sqrt{3+t^2} }{3}  \; \tilde{\nu}_R^{\dagger} \left(\partial^{\mu} \tilde{\nu}_R\right) Z^{\prime}_{\mu} 
   - \frac{1}{18} \sum_j \lambda_j \tilde{\nu}_R^{\dagger} S_j \tilde{\nu}_R ,
   \label{sneutrinoint}
    \end{equation}
%
  where,
\begin{eqnarray}\lambda_j = g \left( a_{1j} \left(-3+2 t^2 \right) v_{\eta_1}
-a_{2j} \left(3+2 t^2\right) v_{\eta'_{1}}-
 a_{3j} \left(3+4 t^2\right)  v_{\rho} 
+ a_{4j}    \left(-3+4 t^2\right)                       v_{\rho' }\right) \   \ 
\end{eqnarray}
  with  $j=1,2,3,4$ and $S_j= H^{\prime}, h^\prime, H, h $, respectively.  The coefficients  $a_{ij}$ are the mixing parameters involving the Higgs.  They are calculated numerically.
 
 Other set of relevant interactions for our calculations are those involving quarks,  leptons and $Z^{\prime}$ provided by the following  lagrangian,
 \begin{equation}
  {\cal L} \supset 
   i\sum_f \bar{f}\gamma^{\mu}  \left( g^{fZ^{\prime}}_ {l.h.} P_L+ g^{fZ^{\prime}}_ {r.h.} P_R \right)f Z^{\prime}_{\mu} +  \sum_{f,j}  \lambda_{f} \bar{f} S_j f
   \label{lagr}
  \end{equation}
 where $j=1,2,3,4$ with $S_j= H^{\prime}, h^\prime, H, h $, respectively and $P_{R,L}=\frac{1}{2}\left(1\pm\gamma_5\right)$. The couplings $g^{fZ^{\prime}}_ {l.h.}$, $g^{fZ^{\prime}}_ {r.h.}$ and $\lambda_{f}$ are given by Table \ref{tab:table1}, the parameters $M_{e,q}$ are the physical masses of charged leptons and quarks, respectively.

 \begin{table}[h!]
  \centering
  \caption{$Z^\prime$ and scalars couplings of Eq. (\ref{lagr}).}
  \label{tab:table1}
  \begin{tabular}{cccc}
   \hline\hline
   
    Fermions($f$) & $g^{fZ^{\prime}}_ {l.h.}$ & $g^{fZ^{\prime}}_ {r.h.}$ & $\lambda_{f}$\\ \hline
    $e_i$ &  $-\frac{g\,\left(3-2t^2\right)}{12 \sqrt{3+t^2}}$ & $\frac{6g\,t^2}{12 \sqrt{3+t^2}}$& $-\frac{2 a_{4j} M_{e_i}}{ v_{\rho^\prime}} $\\
    $\nu_{Ri}$ & $-\frac{g\,\left(3-2t^2\right)}{12 \sqrt{3+t^2}}$ & $0$& $0$\\
    $u,c$ & $\frac{3g}{12 \sqrt{3+t^2}}$ & $-\frac{4g\,t^2}{12 \sqrt{3+t^2}}$& $-\frac{2 a_{3j} M_{u,c}}{ v_{\rho} }$\\ 
    $d,s$ & $\frac{3g}{12 \sqrt{3+t^2}}$ & $\frac{2g\,t^2}{12 \sqrt{3+t^2}}$& $-\frac{2 a_{1j} M_{d,s}} {v_{\eta_1}} $\\
    $b$ & $-\frac{g\,\left(3+2t^2\right)}{12 \sqrt{3+t^2}}$ &  $\frac{2g\,t^2}{12 \sqrt{3+t^2}}$& $-\frac{2 a_{4j} M_b}{v_{\rho^\prime}} $\\
    $t$ & $-\frac{g\,\left(3+2t^2\right)}{12 \sqrt{3+t^2}}$ &   $-\frac{4g\,t^2}{12 \sqrt{3+t^2}}$& $-\frac{2 a_{2j} M_t}{ v_{\eta_1^\prime}} $\\ \hline\hline
    \end{tabular}
\end{table}

\bibliography{sneutrino.bib}
\end{document}